\documentclass[aps,pra,floatfix,showpacs,preprint,superscriptaddress]{revtex4-1}
\usepackage{bm}
\usepackage{mathrsfs}
\usepackage{amsmath}
\usepackage{amssymb}
\usepackage{revsymb}
\usepackage{accents}
\usepackage{graphicx}
\begin{document}
\title{Nonlinear trident pair production in an arbitrary plane wave: a focus on the properties of the transition amplitude}
\author{F. Mackenroth}
\email{mafelix@pks.mpg.de}
\affiliation{Max Planck Institute for the Physics of Complex Systems, N\"{o}thnitzer Stra\ss e 38, D-01187 Dresden,
Germany}
\author{A. \surname{Di Piazza}}
\email{dipiazza@mpi-hd.mpg.de}
\affiliation{Max Planck Institute for Nuclear Physics, Saupfercheckweg 1, D-69117 Heidelberg, Germany}
\date{\today}

\begin{abstract}
The process of nonlinear electron-positron pair production by an electron colliding with an arbitrary plane wave electromagnetic field (nonlinear trident pair production) is studied analytically and numerically. Special emphasis is put on the properties of the transition amplitude. In fact, its original expression as resulting from applying the Wick's theorem turns out to be divergent and not to manifestly fulfill gauge invariance. By restoring the latter, the amplitude is regularized and investigated in different regimes. In particular, the amplitude is divided into a two-step and a one-step contributions, depending on the scaling dependence on the laser pulse duration. The corresponding contributions to the positron angular distribution spectra and the resulting interference terms are studied numerically emphasizing the possibility of measuring experimentally the contribution of the one-step contribution.

  
\end{abstract}
 
\maketitle

\section{Introduction}

The prediction of QED that massive particle-antiparticle pairs can be created solely from electromagnetic fields is certainly among its most thrilling. The recent progress in laser technology (see, e.g., \cite{Yanovsky_etal_2008,VulcanLaser,ELI_WhiteBook,ExtremeLight}) opens the possibility of observing the production of electron-positron pairs in the collision of high-energy photons and intense laser beams (nonlinear Breit-Wheeler pair production (NBWPP)). This process has been thoroughly investigated theoretically in recent years, also accounting for effects of the laser pulse form \cite{Nousch_etal_2012,Krajewska_etal_2013,Gonoskov_etal_2013,Jansen_Mueller_2013,Titov_etal_2014,Augustin_Mueller_2014,Meuren_etal_2016} (see \cite{DiPiazza_etal_2012} for publications until 2012). Since usually high-energy photons are produced via electron back-scattering, alternatively, an electron-positron pair can be produced inside a strong laser field by a high-energy photon emitted by an ultrarelativistic electron colliding with the same laser field (nonlinear trident pair production (NTPP)). Conventionally, the process can be described as commencing via two channels where loosely speaking the photon emission and pair production either occur at the same laser phase (\textit{direct} channel) or at two separate ones (\textit{cascade} channel). A unified theoretical description based on strong-field QED has been recently proposed to analyze NTPP in a plane wave \cite{Hu_Mueller_2010,Ilderton_2011} and in a constant-crossed field \cite{King_Ruhl_2013}.

For an electron (mass $m$ and charge $e<0$) of initial four-momentum $p_i^\mu = (\varepsilon_i,\bm{p}_i)$, with $\varepsilon_i=\sqrt{m^2+p_i^2}$ (units with $\hbar=c=4\pi\epsilon_0=1$ are used throughout), colliding with a plane wave of central angular frequency $\omega_0$, electric field amplitude $E$ and central wave four-vector $k_0^\mu = (\omega_0,\bm{k}_0)$, the total NTPP probability is controlled by the two Lorentz- and gauge-invariant parameters $\xi = |e|E/m\omega_0$ and $\chi = ((k_0p_i)/m^2) E/E_\text{cr}$, where $E_\text{cr} = m^2 /\left|e\right|$ is the critical field of QED \cite{Ritus1985} and we introduced the short notation $(ab) = a_\mu b^\mu$ (the metric tensor is $\eta^{\mu\nu}=\text{diag}(+1,-1,-1,-1)$). An exact inclusion of the laser field in the calculations is necessary for $\xi\gtrsim 1$ \cite{Furry_1951,LandauIV}, which is nowadays routinely achieved at optical laser facilities \cite{VulcanLaser}. At $\xi\gg 1$ NTPP occurs with the absorption of a large number of laser photons and is essentially controlled only by the parameter $\chi$: it is exponentially suppressed for $\chi\ll 1$ and becomes sizable at $\chi\gtrsim 1$. Present day technology allows for optical lasers with $\xi\sim 10^2$ \cite{Yanovsky_etal_2008} and for electron beams with $\varepsilon_i\approx 4\;\text{GeV}$ produced via laser wake-field acceleration \cite{Leemans_2014}, allowing for a thorough experimental investigation of NTPP also within all-optical setups.

So far only one experiment on NTPP has been successfully carried out \cite{Bula_etal_1996,Bamber_etal_1999}, where, however, the direct channel was strongly suppressed. For some time quantitative theoretical studies of NTPP were available only in the idealized cases of monochromatic lasers \cite{Hu_Mueller_2010} and constant-crossed fields \cite{King_Ruhl_2013}. The latter study particularly assessed the validity of approximating trident pair production as a sequence of photon emission and Breit-Wheeler pair production, i.e., neglecting the direct channel, in a constant-crossed field background, as is commonly done in numerical simulations of laser-plasma interactions \cite{Gonoskov_etal_2015,Ridgers_etal_2014,Erber_1966}, and neglected the interference between exchange diagrams. The possibility of suppressing the cascade channel has also been discussed in \cite{Hu_Mueller_2010} in the perturbative regime $\xi\ll 1$ at $\omega_0\sim 10\;\text{eV}$. Relying on the laser field being (almost) monochromatic, in that regime the direct channel is found to be either dominating or comparable with respect to the cascade channel for different laser frequencies. Most recently, however, studies of NTPP in a constant crossed field were amended to also include the exchange diagram interference \cite{King_Fedotov_2018}, finding it to further suppress the contribution of the direct channel for small quantum parameters $\chi$. Additionally, an investigation of the full probability of NTPP in an arbitrary plane wave field revealed that for large $\xi$, the direct contribution as well as the interference between the direct and cascade amplitudes are negligible \cite{Dinu_Torgrimsson_2018}, further supporting the basic assumption underlying numerical particle-in-cell schemes to approximate NTPP as the product of the probabilities for photon emission and pair production. Previous studies had mostly focused on the cascade contribution, which was obtained by employing the optical theorem and the two-loop mass operator \cite{Baier_1972,Ritus_1972,Morozov_1975,Baier_etal_1991}.

The present work focuses on different formal aspects of the full NTPP amplitude in arbitrary plane wave fields. We put forward a scheme to analytically disentangle the direct and the cascade channels of NTPP in an arbitrary plane wave field putting particular emphasis on the amplitude's gauge invariance. We demonstrate how this disentanglement can be employed to identify an explicit cascade contribution, facilitating to identify the remaining parts as true second-order non-cascade contribution. We also show how on amplitude level this split-up naturally yields a phase-ordered cascade and a direct contribution depending only on one laser phase variable. Concerning the experimental observability, we indicate by means of numerical simulations how in ultra-short laser pulses with $\xi\gg 1$ the two channels scale in the energy distribution of the produced positron and one of the electrons.

\section{The transition amplitude}

\begin{figure}[t]
 \centering
 \includegraphics[width=\linewidth]{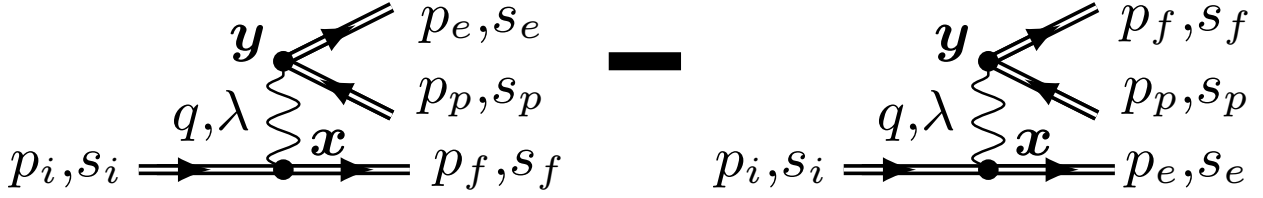}
 \caption{Lowest-order Feynman diagrams of NTPP.}
 \label{fig:NTPP_FeynDiag}
\end{figure}

The background plane-wave field is described by the four-vector potential $A^{\mu}(\phi)$, which only depends on the light-cone time $\phi=t-\bm{n}\cdot \bm{x}$. Here, the unit vector $\bm{n}$ indicates the propagation direction of the plane wave. By introducing the four-dimensional quantity $n^{\mu}=(1,\bm{n})$ and by recalling that the metric tensor reads $\eta^{\mu\nu}=\text{diag}(+1,-1,-1,-1)$, it is $\phi=(nx)$. Having in mind obvious differential properties of the four-vector potential $A^{\mu}(\phi)$ and its derivatives, it is clear that $A^{\mu}(\phi)$ is a solution of the free Maxwell's equation $\square A^{\mu}=0$, with $\square=\partial_{\nu}\partial^{\nu}$. We also assume to work in the Lorenz-gauge $\partial_{\mu}A^{\mu}=0$, with the additional constraint $A^0(\phi)=0$. Thus, if we represent $A^{\mu}(\phi)$ in the form $A^{\mu}(\phi)=(0,\bm{A}(\phi))$, then the Lorenz-gauge condition implies $\bm{n}\cdot\bm{A}'(\phi)=0$, with the prime indicating the derivative with respect to $\phi$. If we make the additional physically reasonable assumption that $\lim_{\phi\to \pm\infty}\bm{A}(\phi)=0$, it results that $\bm{n}\cdot\bm{A}(\phi)=0$. By also introducing two four-vectors $a_j^{\mu}=(0,\bm{a}_j)$, with $j=1,2$, such that $(na_j)=-\bm{n}\cdot\bm{a}_j=0$ and $(a_ia_j)=-\bm{a}_i\cdot\bm{a}_j=-\delta_{ij}$, the most general form of the vector potential $\bm{A}(\phi)$ reads $\bm{A}(\phi)=\psi_1(\phi)\bm{a}_1+\psi_2(\phi)\bm{a}_2$, where the two functions $\psi_j(\phi)$ are arbitrary provided that $\lim_{\phi\to \pm\infty}\psi_j(\phi)=0$ and that they are analytically sufficiently well-behaved. By introducing $\tilde{n}^{\mu}=(1,-\bm{n})/2$, it is clear that the four-dimensional quantities $n^{\mu}$, $\tilde{n}^{\mu}$, and $a^{\mu}_j$ fulfill the completeness relation: $\eta^{\mu\nu}=n^{\mu}\tilde{n}^{\nu}+\tilde{n}^{\mu}n^{\nu}-a_1^{\mu}a_1^{\nu}-a_2^{\mu}a_2^{\nu}$ (note that $(n\tilde{n})=1$ and $(\tilde{n}a_j)=0$). Below, we will refer to the longitudinal ($n$) direction as the direction along the unit vector $\bm{n}$ and to the transverse ($\perp$) plane as the plane spanned by the two perpendicular unit vectors $\bm{a}_j$. In this respect, together with the light-cone time $\phi=t-x_n$, with $x_n=\bm{n}\cdot \bm{x}$, we also introduce the remaining three light-cone coordinates $T=(\tilde{n}x)=(t+x_n)/2$, and $\bm{x}_{\perp}=(x_{a_1},x_{a_2})=(\bm{x}\cdot\bm{a}_1,\bm{x}\cdot\bm{a}_2)$. Analogously, the light-cone coordinates of an arbitrary four-vector $v^{\mu}=(v^0,\bm{v})$ will be indicated as $v_-=(nv)=v^0-v_n$, with $v_n=\bm{n}\cdot \bm{v}$, $v_+=(\tilde{n}v)=(v^0+v_n)/2$, and $\bm{v}_{\perp}=(v_{a_1},v_{a_2})=(\bm{v}\cdot\bm{a}_1,\bm{v}\cdot\bm{a}_2)$.

Now, the amplitude $S_{fi}$ of nonlinear trident pair production in the Furry picture at the leading order is given by (see~Fig. 1 for the corresponding Feynman diagrams):
\begin{equation}
S_{fi}=ie^2\int d^4xd^4y\,\bar{U}_{p_e,s_e}(y)\gamma^{\mu}V_{p_p,s_p}(y)D_{\mu\nu}(y-x)\bar{U}_{p_f,s_f}(x)\gamma^{\nu}U_{p_i,s_i}(x)-\{f\leftrightarrow e\},
\end{equation}
where $\gamma^{\mu}$ are the Dirac gamma-matrices and
\begin{align}
\label{U}
U_{p,s}(x)&=\bigg[1+\frac{e\hat{n}\hat{A}(\phi)}{2p_-}\bigg]e^{i\left\{-(px)-\int_0^{\phi}d\varphi\left[\frac{e(pA(\varphi))}{p_-}-\frac{e^2A^2(\varphi)}{2p_-}\right]\right\}}\frac{u_{p,s}}{\sqrt{2\varepsilon}},\\
\label{V}
V_{p,s}(x)&=\bigg[1-\frac{e\hat{n}\hat{A}(\phi)}{2p_-}\bigg]e^{i\left\{(px)-\int_0^{\phi}d\varphi\left[\frac{e(pA(\varphi))}{p_-}+\frac{e^2A^2(\varphi)}{2p_-}\right]\right\}}\frac{v_{p,s}}{\sqrt{2\varepsilon}}.
\end{align}
are the positive- and negative-energy Volkov states with on-shell four-momentum $p^{\mu}=(\varepsilon,\bm{p})$ and spin quantum number $s$ \cite{LandauIV}, where
\begin{equation}
D^{\mu\nu}(y-x)=\int\frac{d^4q}{(2\pi)^2}\frac{4\pi}{q^2+i0}\eta^{\mu\nu}e^{-i(q(y-x))}
\end{equation}
is the photon propagator, and where the symbol $\{f\leftrightarrow e\}$ indicates that the expression on its left has to be subtracted with the quantum numbers $p_f,s_f$ and $p_e,s_e$ exchanged according to the Fermi-Dirac statistics. In the above expression of the Volkov states, we have assumed a unit quantization volume, we have introduced the notation $\hat{v}=\gamma^{\mu}v_{\mu}$ for an arbitrary four-dimensional quantity $v_{\mu}$ and the free bi-spinors $u_{p,s}$ and $v_{ps}$, which are solutions of the equations $(\hat{p}-m)u_{p,s}=0$ and $(\hat{p}+m)v_{p,s}=0$, respectively \cite{LandauIV}.

By indicating with an index $x$ and $y$ the light-cone coordinates corresponding to each vertex, due to the symmetry of the plane wave, it is possible to carry out the six integrations on the transverse coordinates and on $T_x$ and $T_y$, such that the amplitude $S_{fi}$ has the form
\begin{equation}
\label{S_fi_i}
\begin{split}
S_{fi}=&\frac{ie^2}{k_-}\frac{2\pi}{\sqrt{16\varepsilon_i\varepsilon_f\varepsilon_e\varepsilon_p}}(2\pi)^3\delta(p_{e,-}+p_{f,-}+p_{p,-}-p_{i,-})\delta^{(2)}(\bm{p}_{e,\perp}+\bm{p}_{f,\perp}+\bm{p}_{p,\perp}-\bm{p}_{i,\perp})\\
&\times\int d\phi_xd\phi_y\int\frac{dq_+}{2\pi}e^{-i[S_{p_e}(\phi_y)-S_{-p_p}(\phi_y)+S_{p_f}(\phi_x)-S_{p_i}(\phi_x)]}\frac{e^{-iq_+(\phi_y-\phi_x)}}{q_+-\frac{\bm{k}^2_{\perp}}{2k_-}+i0}\\
&\times\bar{u}_e\bigg[1-\frac{e\hat{n}\hat{A}(\phi_y)}{2p_{e,-}}\bigg]\gamma^{\mu}\bigg[1-\frac{e\hat{n}\hat{A}(\phi_y)}{2p_{p,-}}\bigg]v_p\bar{u}_f\bigg[1-\frac{e\hat{n}\hat{A}(\phi_x)}{2p_{f,-}}\bigg]\gamma_{\mu}\bigg[1+\frac{e\hat{n}\hat{A}(\phi_x)}{2p_{i,-}}\bigg]u_i\\
&-\{f\leftrightarrow e\},
\end{split}
\end{equation}
where
\begin{equation}
S_p(\phi)=-p_+\phi-\int_0^{\phi}d\varphi\left[\frac{e(pA(\varphi))}{p_-}-\frac{e^2A^2(\varphi)}{2p_-}\right]
\end{equation}
where the single index in the bi-spinors indicates both the corresponding four-momentum and the spin quantum numbers, and where $k_-=p_{i,-}-p_{f,-}=p_{e,-}+p_{p,-}$ and $\bm{k}_{\perp}=\bm{p}_{i,\perp}-\bm{p}_{f,\perp}=\bm{p}_{e,\perp}+\bm{p}_{p,\perp}$. The integral in $q_+$ can also be easily taken by means of the residue method and the amplitude becomes
\begin{equation}
\label{S_fi_f}
\begin{split}
S_{fi}=&\frac{e^2}{\sqrt{16\varepsilon_i\varepsilon_f\varepsilon_e\varepsilon_p}}\frac{2\pi}{k_-}(2\pi)^3\delta(p_{e,-}+p_{f,-}+p_{p,-}-p_{i,-})\delta^{(2)}(\bm{p}_{e,\perp}+\bm{p}_{f,\perp}+\bm{p}_{p,\perp}-\bm{p}_{i,\perp})\\
&\times\int d\phi_xd\phi_y\,\theta(\phi_y-\phi_x)\bar{u}_e\bigg[1-\frac{e\hat{n}\hat{A}(\phi_y)}{2p_{e,-}}\bigg]\gamma^{\mu}\bigg[1-\frac{e\hat{n}\hat{A}(\phi_y)}{2p_{p,-}}\bigg]v_p\\
&\times\bar{u}_f\bigg[1-\frac{e\hat{n}\hat{A}(\phi_x)}{2p_{f,-}}\bigg]\gamma_{\mu}\bigg[1+\frac{e\hat{n}\hat{A}(\phi_x)}{2p_{i,-}}\bigg]u_ie^{-i[S_{BW}(\phi_y)+S_{C}(\phi_x)]}-\{f\leftrightarrow e\},
\end{split}
\end{equation}
where $\theta(\cdot)$ is the step function and
\begin{align}
\begin{split}
S_{C}(\phi)=&-k_+\phi+S_{p_f}(\phi)-S_{p_i}(\phi)\\
=&\int_0^{\phi}d\varphi\left[p_{i,+}-p_{f,+}-k_++\frac{e(p_iA)}{p_{i,-}}-\frac{e(p_fA)}{p_{f,-}}-\frac{e^2A^2}{2p_{i,-}}+\frac{e^2A^2}{2p_{f,-}}\right]
\end{split}\\
\begin{split}
S_{BW}(\phi)=&k_+\phi+S_{p_e}(\phi)-S_{-p_p}(\phi)\\
=&\int_0^{\phi}d\varphi\left[k_+-p_{e,+}-p_{p,+}+\frac{e(p_pA)}{p_{p,-}}-\frac{e(p_eA)}{p_{e,-}}+\frac{e^2A^2}{2p_{p,-}}+\frac{e^2A^2}{2p_{e,-}}\right]
\end{split}
\end{align}
are the phases of non-linear Compton scattering and of non-linear Breit-Wheeler pair production, respectively. It is worth noticing that, after performing the integral in $q_+$ the four-momentum of the intermediate photon appears as being on-shell, i.e., $k_+=\bm{k}^2_{\perp}/2k_-$. Below, we will only consider the experimentally most relevant case of a linearly polarized plane wave. Thus, we write the four-vector potential as $A^{\mu}(\phi)=A_0^{\mu}\psi(\phi)$, such that $S_{C}(\phi)=\int_0^{\phi}d\varphi[\alpha_C\psi(\varphi)+\beta_C\psi^2(\varphi)+\gamma_C]$ and $S_{BW}(\phi)=\int_0^{\phi}d\varphi[\alpha_{BW}\psi(\varphi)+\beta_{BW}\psi^2(\varphi)+\gamma_{BW}]$, with
\begin{align}
\alpha_C&=\frac{e(p_iA_0)}{p_{i,-}}-\frac{e(p_fA_0)}{p_{f,-}}, && \alpha_{BW}=\frac{e(p_pA_0)}{p_{p,-}}-\frac{e(p_eA_0)}{p_{e,-}},\\
\beta_C&=\frac{e^2A_0^2}{2}\frac{k_-}{p_{i,-}p_{f,-}}, && \beta_{BW}=\frac{e^2A_0^2}{2}\frac{k_-}{p_{e,-}p_{p,-}},\\
\gamma_C&=p_{i,+}-p_{f,+}-k_+, && \gamma_{BW}=k_+-p_{e,+}-p_{p,+}.
\end{align}
From the expression of the amplitude $S_{fi}$ in Eq. (\ref{S_fi_f}) once could think that the whole amplitude only contains a contribution to the two-step channel or cascade process, as it contains a phase-ordered double integral of the ``product'' of the nonlinear Compton scattering amplitude and of the nonlinear Breit-Wheeler pair production amplitudes, both with the photon being on-shell. This is, however, not the case mainly because the amplitude $S_{fi}$ contains contributions from all polarizations of the intermediate photon, whereas the cascade process only stems from photons having transverse polarization. Another reason is that the amplitude in Eq. (\ref{S_fi_f}) is, on the one hand, still divergent because some terms do not contain the external field in the pre-exponential function and, on the other hand, not manifestly gauge-invariant. The first reason, though, is less fundamental in the sense that, in the case of a constant-crossed field, it does not play a role because also the phase integrals which do not contain the external field in the pre-exponential function do converge. In the present case, precisely the phase integrals corresponding to these terms diverge.

First, we impose that the amplitude $S_{fi}$ is manifestly gauge invariant by going back to the expression in Eq. (\ref{S_fi_i}) and by requiring that the amplitude does not change if the tensor $\eta^{\mu\nu}$ in the photon propagator is replaced by $\eta^{\mu\nu}+q^{\mu}\lambda^{\nu}(q)+q^{\nu}\lambda^{\mu}(q)$, where $\lambda^{\mu}(q)$ is an arbitrary function of $q^{\mu}$ and it is clear from the appearance of the energy-momentum conserving delta-functions, that we can already assume here that $q_-=k_-$ and $\bm{q}_{\perp}=\bm{k}_{\perp}$ from the beginning. Moreover, in order to obtain all the required regularization conditions for the integrals, it is sufficient to assume that $\lambda^{\mu}(q)$ is a constant four-vector. Then, since QED is gauge invariant, once the integrals are convergent, we can be confident that the resulting amplitude is gauge-invariant. Now, one can easily show that by exploiting the energy-momentum conservation laws: $p_i^{\mu}+(k_++p_{f,+}-p_{i,+})n^{\mu}=p_f^{\mu}+k^{\mu}$ and $k^{\mu}+(p_{e,+}+p_{p,+}-k_+)n^{\mu}=p_e^{\mu}+p_p^{\mu}$, the invariance of the amplitude under the mentioned gauge transformation is guaranteed if the ``regularizing'' conditions
\begin{align}
\label{f_a}
\begin{split}
&\int d\phi\,\psi^a(\phi)e^{-i[S_C(\phi)+S_{BW}(\phi)]}\\
&=-i\int d\phi_x d\phi_y\,\theta(\phi_y-\phi_x)[\alpha_C\psi(\phi_x)+\beta_C\psi^2(\phi_x)+\gamma_C]\psi^a(\phi_y)e^{-i[S_C(\phi_x)+S_{BW}(\phi_y)]},
\end{split}\\
\label{f_b}
\begin{split}
&\int d\phi\,\psi^b(\phi)e^{-i[S_C(\phi)+S_{BW}(\phi)]}\\
&=i\int d\phi_x d\phi_y\,\theta(\phi_y-\phi_x)[\alpha_{BW}\psi(\phi_y)+\beta_{BW}\psi^2(\phi_y)+\gamma_{BW}]\psi^b(\phi_x)e^{-i[S_C(\phi_x)+S_{BW}(\phi_y)]},
\end{split}
\end{align}
with $a,b=0,1,2$ are fulfilled. By introducing the quantities
\begin{align}
f_a=&\int d\phi\,\psi^a(\phi)e^{-i[S_C(\phi)+S_{BW}(\phi)]},\\
f_{ab}=&\int d\phi_x d\phi_y\,\theta(\phi_y-\phi_x)\psi^a(\phi_x)\psi^b(\phi_y)e^{-i[S_C(\phi_x)+S_{BW}(\phi_y)]},
\end{align}
it is clear that, among all of them, only $f_0$, $f_{00}$, $f_{01}$, $f_{10}$, $f_{02}$, and $f_{20}$ need to be regularized. The conditions in Eqs. (\ref{f_a},\ref{f_b}) already guarantee that
\begin{align}
f_{0j}&=\frac{1}{\gamma_C}(if_j-\alpha_Cf_{1j}-\beta_Cf_{2j}),\\
f_{j0}&=-\frac{1}{\gamma_{BW}}(if_j+\alpha_{BW}f_{j1}+\beta_{BW}f_{j2}),
\end{align}
with $j=1,2$. By subtracting now Eq. (\ref{f_a}) and Eq. (\ref{f_b}), we obtain that $f_{00}=-(\alpha_Cf_{10}+\beta_Cf_{20}+\alpha_{BW}f_{01}+\beta_{BW}f_{02})/(\gamma_C+\gamma_{BW})$ and, by exploiting the above regularizing relations for $f_{0j}$ and $f_{j0}$, that
\begin{equation}
\begin{split}
f_{00}=&\frac{1}{\gamma_C\gamma_{BW}}\left[i\frac{\alpha_C\gamma_C-\alpha_{BW}\gamma_{BW}}{\gamma_C+\gamma_{BW}}f_1+i\frac{\beta_C\gamma_C-\beta_{BW}\gamma_{BW}}{\gamma_C+\gamma_{BW}}f_2\right.\\
&+\alpha_C\alpha_{BW}f_{11}+\beta_C\beta_{BW}f_{22}+\alpha_C\beta_{BW}f_{12}+\alpha_{BW}\beta_Cf_{21}\bigg].
\end{split}
\end{equation}
Finally, the regularization condition for $f_0$ is obtained by summing Eq. (\ref{f_a}) and Eq. (\ref{f_b}) as we obtain $f_0=-(i/2)[(\gamma_C-\gamma_{BW})f_{00}+\alpha_Cf_{10}-\alpha_{BW}f_{01}+\beta_Cf_{20}-\beta_{BW}f_{02}]$ and then, after some algebra,
\begin{equation}
f_0=-\frac{1}{\gamma_C+\gamma_{BW}}[(\alpha_C+\alpha_{BW})f_1+(\beta_C+\beta_{BW})f_2].
\end{equation}

Now that all integrals are regularized, we can appropriately replace the divergent integrals in Eq. (\ref{S_fi_f}) and the regularized, explicitly gauge-invariant amplitude reads
\begin{equation}
\label{S_fi_gi}
\begin{split}
S_{fi}=&\frac{e^2}{\sqrt{16\varepsilon_i\varepsilon_f\varepsilon_e\varepsilon_p}}\frac{2\pi}{k_-}(2\pi)^3\delta(p_{e,-}+p_{f,-}+p_{p,-}-p_{i,-})\delta^{(2)}(\bm{p}_{e,\perp}+\bm{p}_{f,\perp}+\bm{p}_{p,\perp}-\bm{p}_{i,\perp})\\
&\times\bigg\langle i\int d\phi\,e^{-i[S_{BW}(\phi)+S_{C}(\phi)]}\bigg\{\frac{\bar{u}_e\gamma^{\mu}v_p\bar{u}_f\gamma_{\mu}u_i}{\gamma_C\gamma_{BW}(\gamma_C+\gamma_{BW})}[(\alpha_C\gamma_C-\alpha_{BW}\gamma_{BW})\psi(\phi)\\
&\qquad\qquad\qquad\qquad\qquad\qquad+(\beta_C\gamma_C-\beta_{BW}\gamma_{BW})\psi^2(\phi)]\\
&\qquad\qquad-\frac{\bar{u}_f\gamma_{\mu}u_i}{2\gamma_C}\bar{u}_e\bigg[\frac{e\hat{n}\hat{A}(\phi)}{p_{e,-}}\gamma^{\mu}+\gamma^{\mu}\frac{e\hat{n}\hat{A}(\phi)}{p_{p,-}}-\frac{e^2A^2(\phi)\hat{n}}{p_{e,-}p_{p,-}}n^{\mu}\bigg]v_p\\
&\qquad\qquad+\frac{\bar{u}_e\gamma_{\mu}v_p}{2\gamma_{BW}}\bar{u}_f\bigg[\frac{e\hat{n}\hat{A}(\phi)}{p_{f,-}}\gamma^{\mu}-\gamma^{\mu}\frac{e\hat{n}\hat{A}(\phi)}{p_{i,-}}+\frac{e^2A^2(\phi)\hat{n}}{p_{f,-}p_{i,-}}n^{\mu}\bigg]u_i\bigg\}\\
&+\int d\phi_xd\phi_y\,\theta(\phi_y-\phi_x)M_{BW}^{\mu}(\phi_y)M_{C,\mu}(\phi_x)\bigg\rangle-\{f\leftrightarrow e\},
\end{split}
\end{equation}
where we have introduced the regularized integrands
\begin{align}
\label{M_C_R}
\begin{split}
M_C^{\mu}(\phi)&=\bar{u}_f\bigg\{-\frac{\gamma^{\mu}}{\gamma_C}[\alpha_C\psi(\phi)+\beta_C\psi^2(\phi)]-\frac{e\hat{n}\hat{A}(\phi)}{2p_{f,-}}\gamma^{\mu}+\gamma^{\mu}\frac{e\hat{n}\hat{A}(\phi)}{2p_{i,-}}\\
&\qquad-\frac{e^2A^2(\phi)\hat{n}}{2p_{f,-}p_{i,-}}n^{\mu}\bigg\}u_ie^{-iS_C(\phi)},
\end{split}\\
\label{M_BW_R}
\begin{split}
M_{BW}^{\mu}(\phi)&=\bar{u}_e\bigg\{-\frac{\gamma^{\mu}}{\gamma_{BW}}[\alpha_{BW}\psi(\phi)+\beta_{BW}\psi^2(\phi)]-\frac{e\hat{n}\hat{A}(\phi)}{2p_{e,-}}\gamma^{\mu}-\gamma^{\mu}\frac{e\hat{n}\hat{A}(\phi)}{2p_{p,-}}\\
&\qquad+\frac{e^2A^2(\phi)\hat{n}}{2p_{e,-}p_{p,-}}n^{\mu}\bigg\}v_pe^{-iS_{BW}(\phi)}
\end{split}
\end{align}
of the amplitudes of nonlinear Compton scattering and nonlinear Breit-Wheeler pair production, respectively [it is clear that the substitution $e\leftrightarrow f$ has to be carried out also inside these amplitudes in Eq. (\ref{S_fi_gi})]. The gauge-invariant expression (\ref{S_fi_gi}) of the amplitude $S_{fi}$ is already close to the separation between one-step (direct) channel and two-step (cascade) channel that we want to obtain. As we have already mentioned, we still need to isolate in the cascade amplitude only the contribution due to the two transverse polarizations of the intermediate photon. It is convenient to construct a light-cone basis with the light-like quantities $k^{\mu}$ and $n^{\mu}$, and with the two transverse polarization four-vectors $\Lambda_j^{\mu}=(n^{\mu}a_j^{\nu}-n^{\nu}a_j^{\mu})k_{\nu}/k_-$. In fact, all these quantities fulfill the completeness relation $\eta^{\mu\nu}=(n^{\mu}k^{\nu}+n^{\nu}k^{\mu})/k_--\Lambda_1^{\mu}\Lambda_1^{\nu}-\Lambda_2^{\mu}\Lambda_2^{\nu}$ and we can replace this expression of $\eta^{\mu\nu}$ in all the Lorentz contractions in Eq. (\ref{S_fi_gi}). The result is
\begin{equation}
\label{S_fi_final}
\begin{split}
S_{fi}=&\frac{e^2}{\sqrt{16\varepsilon_i\varepsilon_f\varepsilon_e\varepsilon_p}}\frac{2\pi}{k_-}(2\pi)^3\delta(p_{e,-}+p_{f,-}+p_{p,-}-p_{i,-})\delta^{(2)}(\bm{p}_{e,\perp}+\bm{p}_{f,\perp}+\bm{p}_{p,\perp}-\bm{p}_{i,\perp})\\
&\times\bigg\langle i\int d\phi\,e^{-i[S_{BW}(\phi)+S_{C}(\phi)]}\bigg\{\frac{2}{k_-}\frac{\bar{u}_e\hat{n}v_p\bar{u}_f\hat{n}u_i}{\gamma_C+\gamma_{BW}}[(\alpha_C+\alpha_{BW})\psi(\phi)+(\beta_C+\beta_{BW})\psi^2(\phi)]\\
&-\sum_j\frac{\bar{u}_e\hat{\Lambda}_jv_p\bar{u}_f\hat{\Lambda}_ju_i}{\gamma_C\gamma_{BW}(\gamma_C+\gamma_{BW})}[(\alpha_C\gamma_C-\alpha_{BW}\gamma_{BW})\psi(\phi)+(\beta_C\gamma_C-\beta_{BW}\gamma_{BW})\psi^2(\phi)]\\
&+\sum_j\frac{\bar{u}_f\hat{\Lambda}_ju_i}{2\gamma_C}\bar{u}_e\bigg[\frac{e\hat{n}\hat{A}(\phi)\hat{\Lambda}_j}{p_{e,-}}+\frac{e\hat{\Lambda}_j\hat{n}\hat{A}(\phi)}{p_{p,-}}\bigg]v_p-\frac{\bar{u}_e\hat{\Lambda}_jv_p}{2\gamma_{BW}}\bar{u}_f\bigg[\frac{e\hat{n}\hat{A}(\phi)\hat{\Lambda}_j}{p_{f,-}}-\frac{e\hat{\Lambda}_j\hat{n}\hat{A}(\phi)}{p_{i,-}}\bigg]u_i\bigg\}\\
&-\sum_j\int d\phi_xd\phi_y\,\theta(\phi_y-\phi_x)[M_{BW,\mu}(\phi_y)\Lambda_j^{\mu}][M_{C,\nu}(\phi_x)\Lambda_j^{\nu}]\bigg\rangle-\{f\leftrightarrow e\},
\end{split}
\end{equation}
It should be noticed that the gauge invariance of the amplitude $S_{fi}$ does not imply that the contribution of the terms proportional to $n^{\mu}k^{\nu}$ and to $n^{\nu}k^{\mu}$ vanishes. In fact, the requirement of gauge invariance has been already exploited and is related to the intermediate photon with four-momentum $q^{\mu}$. Indeed, one can show that if one first constructs a basis with the four quantities $q^{\mu}$, $n^{\mu}$ and $\Lambda_j^{\prime\,\mu}=(n^{\mu}a_j^{\nu}-n^{\nu}a_j^{\mu})q_{\nu}/q_-$, separates out the transverse polarization contribution (notice that after the integrals over the transverse and the $T$ coordinates are taken, one obtains $\Lambda_j^{\prime\,\mu}=\Lambda_j^{\mu}$), and then imposes gauge-invariance, one again obtains Eq. (\ref{S_fi_final}).

The result in Eq. (\ref{S_fi_final}) is our main analytical result. By introducing the reduced amplitudes for the direct and the cascade channels as
\begin{align}
\label{M_d}
\begin{split}
M_d=&i\int d\phi\,e^{-i[S_{BW}(\phi)+S_{C}(\phi)]}\bigg\{\frac{2}{k_-}\frac{\bar{u}_e\hat{n}v_p\bar{u}_f\hat{n}u_i}{\gamma_C+\gamma_{BW}}[(\alpha_C+\alpha_{BW})\psi(\phi)+(\beta_C+\beta_{BW})\psi^2(\phi)]\\
&-\sum_j\frac{\bar{u}_e\hat{\Lambda}_jv_p\bar{u}_f\hat{\Lambda}_ju_i}{\gamma_C\gamma_{BW}(\gamma_C+\gamma_{BW})}[(\alpha_C\gamma_C-\alpha_{BW}\gamma_{BW})\psi(\phi)+(\beta_C\gamma_C-\beta_{BW}\gamma_{BW})\psi^2(\phi)]\\
&+\sum_j\frac{\bar{u}_f\hat{\Lambda}_ju_i}{2\gamma_C}\bar{u}_e\bigg[\frac{e\hat{n}\hat{A}(\phi)\hat{\Lambda}_j}{p_{e,-}}+\frac{e\hat{\Lambda}_j\hat{n}\hat{A}(\phi)}{p_{p,-}}\bigg]v_p-\frac{\bar{u}_e\hat{\Lambda}_jv_p}{2\gamma_{BW}}\bar{u}_f\bigg[\frac{e\hat{n}\hat{A}(\phi)\hat{\Lambda}_j}{p_{f,-}}-\frac{e\hat{\Lambda}_j\hat{n}\hat{A}(\phi)}{p_{i,-}}\bigg]u_i\bigg\}\\
&-\{f\leftrightarrow e\},
\end{split}\\
\label{M_c}
M_c&=-\sum_j\int d\phi_xd\phi_y\,\theta(\phi_y-\phi_x)[M_{BW,\mu}(\phi_y)\Lambda_j^{\mu}][M_{C,\nu}(\phi_x)\Lambda_j^{\nu}]-\{f\leftrightarrow e\},
\end{align}
we can write the differential trident probability $dP$ summed/averaged over all final/initial spin quantum number and integrated over the final electrons' momenta as
\begin{equation}
\begin{split}
dP=\frac{\alpha^2\pi^2}{4}\frac{1}{p_{i,-}}\frac{d^3\bm{p}_p}{(2\pi)^3}\frac{1}{2\varepsilon_p}\sum_{s_i,s_f,s_e,s_p}\int\frac{d^3\bm{p}_e}{(2\pi)^3}\frac{1}{2\varepsilon_e}\frac{1}{k^2_-p_{f,-}}\left[|M_d|^2+|M_c|^2+2\text{Re}(M_d^*M_c)\right]
\end{split}
\end{equation}
where we have exploited the three-dimensional delta-function in the amplitude to take the integral in $d^3\bm{p}_f$ and where $\alpha=e^2$ is the fine-structure constant. It is worth pointing out that the probability corresponding to the term $|M_c|^2$ in the integrand should not be identified yet with the cascade probability. The reason is that the quantity $|M_c|^2$ contains interference terms between different (transverse) polarizations of the intermediate photon and interference terms between the two amplitudes differing by the exchange of the quantum numbers of the two final electrons. In order to clearly isolate what we will call the cascade probability, which reduces to the one computed in \cite{Baier_1972,Ritus_1972,Morozov_1975,Baier_etal_1991} in the case of a constant crossed field, we decompose the direct and the cascade amplitudes as
\begin{align}
M_d&=M^{(ef)}_{d,n}+M^{(ef)}_{d,1}+M^{(ef)}_{d,2}-M^{(fe)}_{d,n}-M^{(fe)}_{d,1}-M^{(fe)}_{d,2},\\
M_c&=M^{(ef)}_{c,1}+M^{(ef)}_{c,2}-M^{(fe)}_{c,1}-M^{(fe)}_{c,2},
\end{align}
with the definition of each single term being clear from the expression in Eqs. (\ref{M_d}) and (\ref{M_c}) [for the sake of clarity we specify that the term $M^{(ef)}_{d,n}$ corresponds to the second line in Eq. (\ref{S_fi_final}) and that the indexes $1$, and $2$ refer to the different transverse polarizations of the intermediate photon]. According to this splitting of the amplitude, we write the differential probability as $dP=dP_c+dP_d+d\mathcal{P}_i$, where
\begin{equation}
\label{dP_c}
\begin{split}
dP_c=&\frac{\alpha^2\pi^2}{4}\frac{1}{p_{i,-}}\frac{d^3\bm{p}_p}{(2\pi)^3}\frac{1}{2\varepsilon_p}\sum_{s_i,s_f,s_e,s_p}\int\frac{d^3\bm{p}_e}{(2\pi)^3}\frac{1}{2\varepsilon_e}\frac{1}{k^2_-p_{f,-}}\\
&\times\left[|M^{(ef)}_{c,1}|^2+|M^{(ef)}_{c,2}|^2+|M^{(fe)}_{c,1}|^2+|M^{(fe)}_{c,2}|^2\right]\\
=&\frac{\alpha^2\pi^2}{2}\frac{1}{p_{i,-}}\frac{d^3\bm{p}_p}{(2\pi)^3}\frac{1}{2\varepsilon_p}\sum_{s_i,s_f,s_e,s_p}\int\frac{d^3\bm{p}_e}{(2\pi)^3}\frac{1}{2\varepsilon_e}\frac{1}{k^2_-p_{f,-}}\left[|M^{(ef)}_{c,1}|^2+|M^{(ef)}_{c,2}|^2\right]
\end{split}
\end{equation}
is the cascade-channel probability, 
\begin{equation}
\begin{split}
dP_d=&\frac{\alpha^2\pi^2}{4}\frac{1}{p_{i,-}}\frac{d^3\bm{p}_p}{(2\pi)^3}\frac{1}{2\varepsilon_p}\sum_{s_i,s_f,s_e,s_p}\int\frac{d^3\bm{p}_e}{(2\pi)^3}\frac{1}{2\varepsilon_e}\frac{1}{k^2_-p_{f,-}}\\
&\times\left[|M^{(ef)}_{d,n}|^2+|M^{(ef)}_{d,1}|^2+|M^{(ef)}_{d,2}|^2+|M^{(fe)}_{d,n}|^2+|M^{(fe)}_{d,1}|^2+|M^{(fe)}_{d,2}|^2\right]\\
=&\frac{\alpha^2\pi^2}{2}\frac{1}{p_{i,-}}\frac{d^3\bm{p}_p}{(2\pi)^3}\frac{1}{2\varepsilon_p}\sum_{s_i,s_f,s_e,s_p}\int\frac{d^3\bm{p}_e}{(2\pi)^3}\frac{1}{2\varepsilon_e}\frac{1}{k^2_-p_{f,-}}\left[|M^{(ef)}_{d,n}|^2+|M^{(ef)}_{d,1}|^2+|M^{(ef)}_{d,2}|^2\right]
\end{split}
\end{equation}
is the direct-channel probability, and $d\mathcal{P}_i=dP-dP_c-dP_d$ is the sum of the several interference terms, which do not need to be reported here (since $d\mathcal{P}_i$ can be negative we have used a different symbol to indicate it). Before passing to the numerical results, we would like to show explicitly how the quantity $P_c=\int dP_c$, with the integral being meant to be over the positron momentum [see Eq. (\ref{dP_c})], reduces to the cascade probability in the local constant field approximation (see \cite{Baier_1972,Ritus_1972,Morozov_1975,Baier_etal_1991}). Since in Eq. (\ref{dP_c}) we decided to write the probability in terms of $M^{(ef)}_{c,j}$, we consider the following one-vertex processes:
\begin{enumerate}
\item nonlinear Compton scattering by an electron with four-momentum $p^{\mu}_i$ and spin quantum number $s_i$ which emits a (real) photon with four-momentum $k^{\mu}$ and (transverse) polarization $j$ and remains with four-momentum $p^{\mu}_f$ and spin quantum number $s_f$;
\item nonlinear Breit-Wheeler pair production by a (real) photon with four-momentum $k^{\mu}$ and (transverse) polarization $j$, which transforms into an electron with four-momentum $p^{\mu}_e$ and spin quantum number $s_e$ and a positron with four-momentum $p^{\mu}_p$ and spin quantum number $s_p$;
\end{enumerate}
The regularized probability amplitudes of these two processes can be written as
\begin{align}
\label{S_C}
S_{C,j}=&-ie\sqrt{\frac{4\pi}{8\varepsilon_i\varepsilon_f\omega}}(2\pi)^3\delta(p_{f,-}+k_--p_{i,-})\delta^{(2)}(\bm{p}_{f,\perp}+\bm{k}_{\perp}-\bm{p}_{i,\perp})\int d\phi_x\,M_{C,\nu}(\phi_x)\Lambda_j^{\nu},\\
\label{S_BW}
S_{BW,j}=&-ie\sqrt{\frac{4\pi}{8\varepsilon_e\varepsilon_p\omega}}(2\pi)^3\delta(p_{e,-}+p_{p,-}-k_-)\delta^{(2)}(\bm{p}_{e,\perp}+\bm{p}_{p,\perp}-\bm{k}_{\perp})\int d\phi_y\,M_{BW,\nu}(\phi_y)\Lambda_j^{\nu},
\end{align}
and we compare these amplitudes with the cascade amplitude
\begin{equation}
\begin{split}
S^{(ef)}_{c,j}=&-e^2\sqrt{\frac{(4\pi)^2}{64\varepsilon_i\varepsilon_f\omega^2\varepsilon_e\varepsilon_p}}\frac{\omega}{k_-}(2\pi)^3\delta(p_{e,-}+p_{f,-}+p_{p,-}-p_{i,-})\delta^{(2)}(\bm{p}_{e,\perp}+\bm{p}_{f,\perp}+\bm{p}_{p,\perp}-\bm{p}_{i,\perp})\\
&\times\int d\phi_xd\phi_y\,\theta(\phi_y-\phi_x)[M_{BW,\mu}(\phi_y)\Lambda_j^{\mu}][M_{C,\nu}(\phi_x)\Lambda_j^{\nu}]
\end{split}
\end{equation}
corresponding to the partial amplitude $M^{(ef)}_{c,j}$. In order to calculate the transition probabilities, we have to square the delta-functions and it is convenient first to use the transformations (we have implicitly employed the last of these transformations already above when we computed the differential probability $dP$)
\begin{align}
\delta(p_{f,-}+k_--p_{i,-})&=\frac{\varepsilon_i}{p_{i,-}}\delta(p_{i,n}-\bar{p}_{i,n}),\\
\delta(p_{e,-}+p_{p,-}-k_-)&=\frac{\omega}{k_-}\delta(k_n-\bar{k}_n),\\
\delta(p_{e,-}+p_{f,-}+p_{p,-}-p_{i,-})&=\frac{\varepsilon_i}{p_{i,-}}\delta(p_{i,n}-\bar{p}'_{i,n})
\end{align}
to the corresponding longitudinal components of the momenta, where the exact expressions of the quantities $\bar{p}_{i,n}$, $\bar{k}_n$, and $\bar{p}'_{i,n}$ are not necessary here. By computing the modulus square of the above amplitudes, we obtain the following probabilities
\begin{align}
\frac{dP_{C,j}}{d\bm{k}^3}&=\frac{e^2}{2}\frac{1}{(2\pi)^3}\sum_{s_i,s_f}\int\frac{d^3\bm{p}_f}{(2\pi)^3}\frac{\omega}{k_-}\frac{\varepsilon_i}{p_{i,-}}(2\pi)^3\delta^{(3)}(\bm{k}-\bm{k}_0)\left\vert\int d\phi_x M_{C,\nu}(\phi_x)\Lambda_j^{\nu}\right\vert^2,\\
P_{BW,j}&=e^2\sum_{s_e,s_p}\int\frac{d^3\bm{p}_p}{(2\pi)^3}\frac{\omega}{k_-}\frac{\varepsilon_e}{p_{e,-}}\left\vert\int d\phi_y M_{BW,\mu}(\phi_y)\Lambda_j^{\mu}\right\vert^2,\\
\label{P_c_ef_j}
\begin{split}
P^{(ef)}_{c,j}&=\frac{e^4}{2}\sum_{s_i,s_f,s_e,s_p}\int\frac{d^3\bm{p}_p}{(2\pi)^3}\int\frac{d^3\bm{p}_f}{(2\pi)^3}\frac{\omega^2}{k_-^2}\frac{\varepsilon_i}{p_{i,-}}\frac{\varepsilon_e}{p_{e,-}}\\
&\times\left\vert\int d\phi_xd\phi_y\,\theta(\phi_y-\phi_x)[M_{BW,\mu}(\phi_y)\Lambda_j^{\mu}][M_{C,\nu}(\phi_x)\Lambda_j^{\nu}]\right\vert^2.
\end{split}
\end{align}
Note that, in order to calculate the cascade probability out of the two probabilities of nonlinear Compton scattering and nonlinear Breit-Wheeler pair production, one needs initially only the differential probability of nonlinear Compton scattering in the photon momentum. In this respect, it was convenient to write the three-dimensional delta-function in terms of the emitted photon momentum and, as above, the expression of the momentum $\bm{k}_0$ is not needed. Also, we observe that it makes physical sense to talk about a cascade process only when the probabilities can be expressed as integrals over laser phases (or times for external fields of different structures) of corresponding probabilities per unit phase (time), which depend only on the local value of the plane wave (external field) at that phase (time). We now first focus on the one-vertex processes and thus imagine to work in the local constant crossed field limit where the classical nonlinearity parameter $\xi$ is very large. When we square the amplitude, for example, of nonlinear Compton scattering, we obtain a double integral in $\phi_x$ and, say, $\phi'_x$ [see Eq. (\ref{S_C})]. Since in the local constant crossed field limit the dominant contribution to the probabilities comes from the region where the quantity $|\phi'_x-\phi_x|$ is much smaller (by a factor of the order of $1/\xi$) than the laser central period \cite{Ritus1985,Baier_b_1998}, it is convenient to pass to the variables $\phi_{x,+}=(\phi'_x+\phi_x)/2$ and $\phi_{x,-}=\phi'_x-\phi_x$ and expand the integrand with respect to $\phi_{x,-}$. The procedure is well known [see, e.g., \cite{DiPiazza_2017}] and it is not necessary to report the details here. It is important to point out that the probability of nonlinear Compton scattering in this limit can be written in the form $dP_{C,j}/d\bm{k}^3=\int d\phi_{x,+} dP_{C,j}(\phi_{x,+})/d\phi_{x,+}d\bm{k}^3$, where
\begin{equation}
\begin{split}
\frac{dP_{C,j}(\phi_{x,+})}{d\phi_{x,+}d\bm{k}^3}=&\frac{e^2}{2}\frac{1}{(2\pi)^3}\sum_{s_i,s_f}\int\frac{d^3\bm{p}_f}{(2\pi)^3}\frac{\omega}{k_-}\frac{\varepsilon_i}{p_{i,-}}(2\pi)^3\delta^{(3)}(\bm{k}-\bm{k}_0)\\
&\times\int d\phi_{x,-}[M_{C,\nu}(\phi_{x,+}-\phi_{x,-}/2)\Lambda_j^{\nu}][M^*_{C,\nu'}(\phi_{x,+}+\phi_{x,-}/2)\Lambda_j^{\nu'}],
\end{split}
\end{equation}
with $dP_{C,j}(\phi_{x,+})/d\phi_{x,+}d\bm{k}^3$ being a non-negative quantity (in this limit) depending only on the plane-wave electromagnetic field calculated at $\phi_{x,+}$. Analogously, one can write in the same limit that $P_{BW,j}=\int d\phi_{y,+} P_{BW,j}(\phi_{y,+})/d\phi_{y,+}$, where
\begin{equation}
\begin{split}
\frac{dP_{BW,j}(\phi_{y,+})}{d\phi_{y,+}}=&e^2\sum_{s_e,s_p}\int\frac{d^3\bm{p}_p}{(2\pi)^3}\frac{\omega}{k_-}\frac{\varepsilon_e}{p_{e,-}}\\
&\times\int d\phi_{y,-}[M_{BW,\mu}(\phi_{y,+}-\phi_{y,-}/2)\Lambda_j^{\mu}][M^*_{BW,\mu'}(\phi_{y,+}+\phi_{y,-}/2)\Lambda_j^{\mu'}],
\end{split}
\end{equation}
with $\phi_{y,+}=(\phi'_y+\phi_y)/2$ and $\phi_{y,-}=\phi'_y-\phi_y$. Now, it is clear that the total cascade probability $\tilde{P}_c$ calculated out of the two elementary processes of nonlinear Compton scattering and nonlinear Breit-Wheeler pair production is given by
\begin{equation}
\label{P_c_tilde}
\begin{split}
\tilde{P}_c&=\sum_j\int d^3\bm{k}\int d\phi_{x,+}d\phi_{y,+}\theta(\phi_{y,+}-\phi_{x,+})\frac{dP_{BW,j}(\phi_{y,+})}{d\phi_{y,+}}\frac{dP_{C,j}(\phi_{x,+})}{d\phi_{x,+}d\bm{k}^3}\\
&=\frac{e^4}{2}\sum_j\sum_{s_i,s_e,s_f,s_p}\int d\phi_{x,+}d\phi_{y,+}\theta(\phi_{y,+}-\phi_{x,+})\int\frac{d^3\bm{p}_p}{(2\pi)^3}\int\frac{d^3\bm{p}_f}{(2\pi)^3}\frac{\omega^2}{k^2_-}\frac{\varepsilon_i}{p_{i,-}}\frac{\varepsilon_e}{p_{e,-}}\\
&\quad\times\int d\phi_{y,-}d\phi_{x,-}[M_{BW,\mu}(\phi_{y,+}-\phi_{y,-}/2)\Lambda_j^{\mu}][M^*_{BW,\mu'}(\phi_{y,+}+\phi_{y,-}/2)\Lambda_j^{\mu'}]\\
&\qquad\qquad\times[M_{C,\nu}(\phi_{x,+}-\phi_{x,-}/2)\Lambda_j^{\nu}][M^*_{C,\nu'}(\phi_{x,+}+\phi_{x,-}/2)\Lambda_j^{\nu'}].
\end{split}
\end{equation}
Now, looking back at Eq. (\ref{P_c_ef_j}) and imagining to perform the sum over $j$ and then work in the local constant field limit, we easily realize that that equation coincides with Eq. (\ref{P_c_tilde}) if the approximation $\theta(\phi_y-\phi_x)\theta(\phi'_y-\phi'_x)\approx\theta(\phi_{y,+}-\phi_{x,+})$ holds in the same limit. In fact, the amplitudes of the elementary processes (nonlinear Compton scattering and nonlinear Breit-Wheeler pair production) are exactly the same [and given by Eqs. (\ref{M_C_R},\ref{M_BW_R})] and the limiting procedure is the same for both equations. The above approximate identity between theta-functions can be proved starting from the identity $\theta(a)\theta(b)=\theta(ab)\theta(a+b)$ valid for any pair of real numbers $a$ and $b$, which in our case provides the identity
\begin{equation}
\theta(\phi_y-\phi_x)\theta(\phi'_y-\phi'_x)=\theta(\phi_{y,+}-\phi_{x,+})\theta((\phi_y-\phi_x)(\phi'_y-\phi'_x)).
\end{equation}
Now, we use the identity $\theta(ab)=1-[\theta(a)-\theta(b)]^2=1-\theta(|a-b|/2-(a+b)/2)$ to finally obtain
\begin{equation}
\label{thetas}
\theta(\phi_y-\phi_x)\theta(\phi'_y-\phi'_x)=\theta(\phi_{y,+}-\phi_{x,+})\left[1-\theta\left(\frac{|\phi_{y,-}-\phi_{x,-}|}{2}-(\phi_{y,+}-\phi_{x,+})\right)\right],
\end{equation}
which approximately turns into the needed equality once we observe that in the local constant field limit we can neglect the small quantity $|\phi_{y,-}-\phi_{x,-}|/2$. One can alternatively define as cascade term only the one coming from the first term in Eq. (\ref{thetas}), which is still exact.

\section{Numerical investigations}

\begin{figure}[t]
\centering
 \includegraphics[width=.5\linewidth]{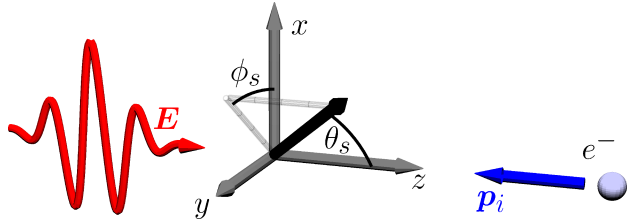}
 \caption{Spherical coordinate frame in which we study the NTPP process with the polar and azimuthal emission angles $(\theta_s,\phi_s)$, respectively, indicated.}
 \label{fig:Figure2}
\end{figure}

In the following we wish to exemplify our analytical findings in a series of numerical test cases. We consider a linearly polarized laser pulse propagating along the $z$-axis, which we additionally assume to collide head-on with an ultrarelativistic electron (see~Fig.~\ref{fig:Figure2}). We then fix the observation directions of the positron and one of the final state electrons and present numerical integrations of the fully differential energy distribution of the NTPP probability for observations in these fixed directions. As we are considering a linearly polarized laser pulse, we expect most of the classical particle dynamics and positron production to occur on the plane identified by the laser propagation direction and the laser polarization direction. In a spherical coordinate frame (see~Fig.~\ref{fig:Figure2}) this plane is denoted by the $x$-$z$ plane, whence we focus most of our discussion on observing particles in either $\phi_s=0$ or $\phi_s=\pi$. Furthermore, we note that in the regime $\varepsilon_i ,\omega \gg m \xi$, as we study here, final state particles in nonlinear Compton scattering and NBWPP are angularly confined around the initial state electron's and photon's propagation direction, respectively, to a narrow cone of opening angle $\theta_s \sim m \xi / \varepsilon_i$ and $\theta_s \sim m \xi / \omega$, respectively. We thus observe the final state particles of NTPP in a direction close to the initial state electron's propagation direction, in a head-on collision along the $z$-axis given by $\theta_s = \pi$. 
\begin{figure}[t]
 \centering
 \includegraphics[width=.5\linewidth]{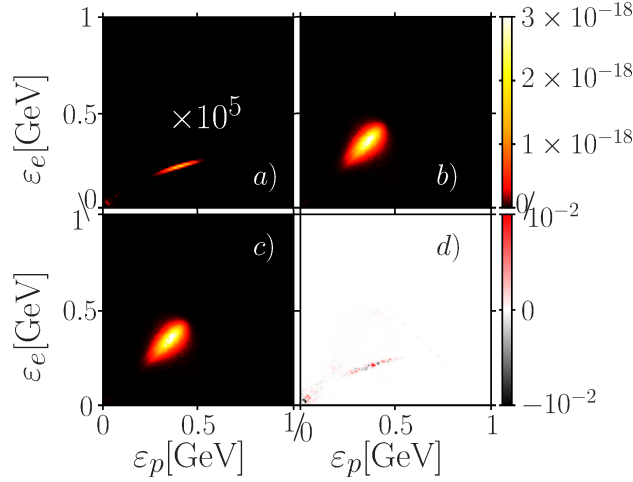}
 \caption{Differential NTPP probability of direct (a), cascade (b) and full (c) channels for the collision of an electron with initial energy $\varepsilon_i = 1$ GeV with a laser pulse of intensity $I=2\times 10^{21}\; \text{W}/\text{cm}^2\ (\xi \approx 22, \chi\approx 0.25)$, the positron observed at $(\theta_s,\phi_s) = (\pi - m\xi/\varepsilon_i,\pi/2)$ and one of the electrons at $(\theta_s,\phi_s) = (\pi - m\xi/\varepsilon_i,0)$. The relative error of the cascade approximation $\mathcal{R}$ is shown in (d).}
 \label{fig:Figure3}
\end{figure}

We begin by studying a case typical of nowadays feasible all-optical experiments in which an electron of initial energy $\varepsilon_i = 1$ GeV collides with a laser pulse of intensity $I=2\times 10^{21}\; \text{W}/\text{cm}^2\ (\xi \approx 22)$, yielding a comparatively small quantum nonlinearity parameter $\chi \approx 0.25$. We find the total emission rate to be completely dominated by the cascade process (see~Fig.~\ref{fig:Figure3} b) and c)), with the direct contribution suppressed by about $5$ orders of magnitude (see~Fig.~\ref{fig:Figure3} a)). The dominance of the cascade contribution is even more obvious from studying the relative error made by approximating the total NTPP probability with the cascade contribution, distinguished by the parameter
\begin{align}
 \mathcal{R} = \frac{dP-dP_c}{dP}.
\end{align}
In the current case we find this parameter to be of percent-level in the particle energy regime where the direct channel is strongest. However, we note that the deviation is not a pure contribution of the direct channel, but an interference effect. Furthermore, we find the probability of NTPP to be centered around the symmetry axis of the energy distribution, indicating that all three final state particles share a comparable amount of energy.

\begin{figure}[t]
 \centering
 \includegraphics[width=.5\linewidth]{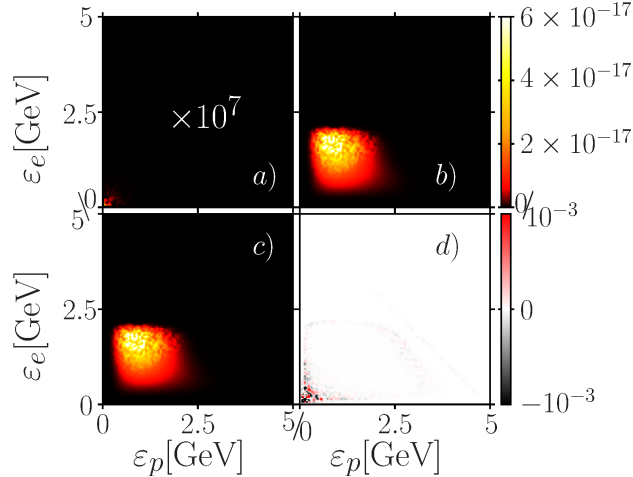}
 \caption{Differential NTPP probability of direct (a), cascade (b) and full (c) channels for the collision of an electron with initial energy $\varepsilon_i = 5$ GeV with a laser pulse of intensity $I=10^{22}\;\text{W}/\text{cm}^2\ (\xi \approx 50, \chi\approx 3)$, the positron observed at $(\theta_s,\phi_s) = (\pi - m\xi/\varepsilon_i,\pi/2)$ and one of the electrons at $(\theta_s,\phi_s) = (\pi - m\xi/\varepsilon_i,0)$. The relative error of the cascade approximation $\mathcal{R}$ is shown in (d).}
 \label{fig:Figure4}
\end{figure}

We continue by studying a case deeper in the nonlinear quantum regime in which an electron of initial energy $\varepsilon_i = 5$ GeV collides with a laser pulse of intensity $I=10^{22}\; \text{W}/\text{cm}^2\ (\xi \approx 50)$, yielding a larger quantum nonlinearity parameter $\chi \approx 3$, likely to be close to the optimum operation parameters for upcoming laser facilities. Due to the increased laser intensity, we find the total emission rate to be even more strongly dominated by the cascade process (see~Fig.~\ref{fig:Figure4} b) and c)), with the direct contribution's suppression increased to $7$ orders of magnitude (see~Fig.~\ref{fig:Figure4} a)) and contributing only at the smallest final state particle energies. The relative error of the cascade approximation is consequently found to be most significant at small final state particle energies, where the direct channel is strongest, but to be overall small on the level of a per mill (s.~fig.~\ref{fig:Figure4} d)).

\begin{figure}[t]
 \centering
 \includegraphics[width=.5\linewidth]{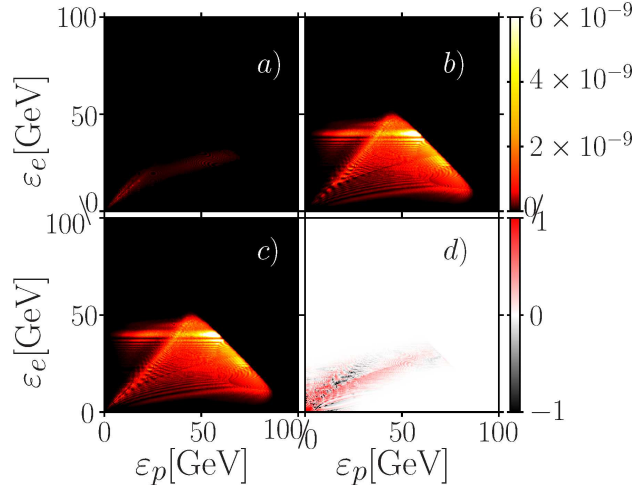}
 \caption{Differential NTPP probability of direct (a), cascade (b) and full (c) channels for the collision of an electron with initial energy $\varepsilon_i = 100$ GeV with a laser pulse of intensity $I=2\times 10^{21}\; \text{W}/\text{cm}^2\ (\xi \approx 22, \chi\approx 26)$, the positron observed at $(\theta_s,\phi_s) = (\pi - m\xi/\varepsilon_i,\pi)$ and one of the electrons at $(\theta_s,\phi_s) = (\pi - m\xi/\varepsilon_i,0)$. The relative error of the cascade approximation $\mathcal{R}$ is shown in (d).}
 \label{fig:Figure5}
\end{figure}

These findings further corroborate the common approximation of higher order nonlinear QED effects, notably NTPP, by their incoherent contributions \cite{Ridgers_etal_2014,Gonoskov_etal_2015}. In order to explore the limitations of this approximation, we turn to a parameter regime where its applicability is expected to be less justified. We study a regime with a high quantum nonlinearity parameter but relatively small laser intensity \cite{King_Ruhl_2013,King_Fedotov_2018}. Consequently, we consider the initial electron to have a very high energy of $\varepsilon_i = 100$ GeV. In combination with a laser intensity of $I=2\times 10^{21}\;\text{W}/\text{cm}^2\ (\xi \approx 22)$ this results in a quantum nonlinearity parameter of $\chi \approx 26$. Analyzing now the direct contribution on the same scale as the cascade and full contributions, we find its impact to be no longer negligible at low particle energies (see~Fig.~\ref{fig:Figure5}). Furthermore, we find larger positron than electron energies to be favored in this regime, as apparent from the asymmetric distributions of the energy spectra (see~Fig.~\ref{fig:Figure5} a), b), c)). Interestingly, in the cascade contribution we also find considerable interference fringes, depending almost exclusively on the electron's energy (see~Fig.~\ref{fig:Figure5} b)). This is most probably due to the fact that the final state contains two indistinguishable electrons whose distributions can interfere. The relative error of the cascade approximation, on the other hand, is found to reach the level of $100\%$ for small final state particle energies (see~Fig.~\ref{fig:Figure5} d)), indicating that at these extreme parameters the cascade approximation starts to lose applicability.

\begin{figure}[t]
 \centering
 \includegraphics[width=.5\linewidth]{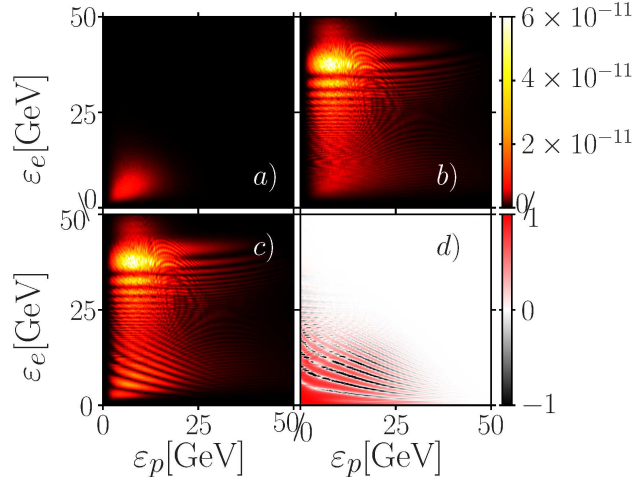}
 \caption{Differential NTPP probability of direct (a), cascade (b) and full (c) channels for the collision of an electron with initial energy $\varepsilon_i = 100$ GeV with a laser pulse of intensity $I=5\times 10^{20}\; \text{W}/\text{cm}^2\ (\xi \approx 11,\chi\approx 13)$, the positron observed at $(\theta_s,\phi_s) = (\pi - m\xi/\varepsilon_i,\pi/2)$ and one of the electrons at $(\theta_s,\phi_s) = (\pi - m\xi/\varepsilon_i,0)$. The relative error of the cascade approximation $\mathcal{R}$ is shown in (d).}
 \label{fig:Figure6}
\end{figure}

We can even further enhance the visibility of the direct channel by considering a lower laser intensity. Furthermore, as a semi-classical picture of NTPP predicts the particle production to be mostly confined to the laser's plane of polarization, we can expect to observe stronger deviations from the cascade model, by observing one of the electrons inside the polarization plane $(\theta_s,\phi_s) = (\pi - m\xi/\varepsilon_i,0)$ but the positron in a direction perpendicularly to this plane $(\theta_s,\phi_s) = (\pi - m\xi/\varepsilon_i,\pi/2)$. We note, however, that for smaller initial electron energies observing the positron perpendicularly to the laser's plane of polarization does not result in a significant contribution from the direct channel (see~Fig.~\ref{fig:Figure3}). For a large initial electron energy of $\varepsilon_i = 100$ GeV, on the other hand, we indeed find that in the collision with a laser pulse of intensity $I=5\times 10^{20}\; \text{W}/\text{cm}^2\ (\xi \approx 11,\chi\approx 13)$, the direct contribution is more pronounced in comparison to the cascade channel (see~Fig.~\ref{fig:Figure6} a)). Again, we find the cascade channel's interference fringes to depend dominantly on the electron's energy (see~Fig.~\ref{fig:Figure6} b)). In the full NTPP probability, however, at small final state particle energies we find the interference fringes to exhibit a dependence on the positron's energy as well. This is a clear indication that the direct channel and interference terms between exchange diagrams start to affect the overall NTPP rate (see~Fig.~\ref{fig:Figure6} c)). Furthermore, we find the total NTPP signal to be significantly enhanced at low particle energies, as is also apparent from the relative error $\mathcal{R}$, which is significant for low particle energies (see~Fig.~\ref{fig:Figure6} d)).

\section{Conclusions}

We have derived a novel splitting of the full scattering matrix amplitude of NTPP in a plane wave of arbitrary shape into a cascade and direct contribution. We found the cascade probability to reduce to the common product of nonlinear Compton scattering and Breit-Wheeler pair production probabilities in the case of a constant crossed field and isolated the contributions of non-cascade parts to NTPP. By squaring the amplitudes we found the observable probabilities for NTPP via the cascade and direct channels and analyzed the latter in exemplary cases. Our numerical analyses further confirmed the applicability of the cascade approximation of NTPP at low initial electron energies and high laser intensities, but also indicated that at high initial electron energies, non-cascade contributions may affect the overall NTPP rate.

\begin{acknowledgments}
The authors acknowledge fruitful discussions with C. H. Keitel, B. King and C. M\"uller.
\end{acknowledgments}

% \bibliography{../Bibliography}

\end{document}